# Direct observation of the transition from indirect to direct bandgap in atomically thin epitaxial MoSe$_2$

Yi Zhang[1,2], Tay-Rong Chang[3], Bo Zhou[1,4,5], Yong-Tao Cui[2,4], Hao Yan[2,4], Zhongkai Liu[2,4], Felix Schmitt[2,4], James Lee[2,4], Rob Moore[2,4], Yulin Chen[1,4,5], Hsin Lin[6], Horng-Tay Jeng[3,7], Sung-Kwan Mo[1,*], Zahid Hussain[1], Arun Bansil[6] and Zhi-Xun Shen[2,4,*]

[1]Advanced Light Source, Lawrence Berkeley National Laboratory, Berkeley, CA 94720, USA

[2]Stanford Institute for Materials and Energy Sciences, SLAC National Accelerator Laboratory, Menlo Park, CA 94025, USA

[3]Department of Physics, National Tsing Hua University, Hsinchu 30013, Taiwan

[4]Geballe Laboratory for Advanced Materials, Departments of Physics and Applied Physics, Stanford University, Stanford, CA 94305, USA

[5]Department of Physics and Clarendon Laboratory, University of Oxford, Parks Road, Oxford, OX1 3PU, UK

[6]Department of Physics, Northeastern University, Boston, MA 02115, USA

[7]Institute of Physics, Academia Sinica, Taipei 11529, Taiwan.

*e-mail: SKMo@lbl.gov; zxshen@stanford.edu

**Quantum systems in confined geometries are host to novel physical phenomena. Examples include quantum Hall systems in semiconductors[1] and Dirac electrons in graphene[2]. Interest in such systems has also been intensified by the recent discovery of a large enhancement in photoluminescence quantum efficiency[3-7] and a potential route to valleytronics[6-8] in atomically thin layers of transition metal dichalcogenides, MX$_2$ (M = Mo, W; X = S, Se, Te), which are closely related to the indirect-to-direct bandgap transition in monolayers[9-12]. Here, we report the first direct observation of the transition from indirect to direct bandgap in monolayer samples by using angle-resolved photoemission spectroscopy on high-quality thin films of MoSe$_2$ with variable thickness, grown by molecular beam epitaxy. The band structure measured experimentally indicates a stronger tendency of monolayer MoSe$_2$ towards a direct bandgap, as well as a larger gap size, than theoretically predicted. Moreover, our finding of a significant spin-splitting of ~180 meV at the valence band maximum of a monolayer MoSe$_2$ film could expand its possible application to spintronic devices.**

The layered transition metal dichalcogenides (TMDs) MX$_2$ (M = Mo, W; X = S, Se, Te), a class of graphene-like two-dimensional materials, have attracted significant interest because they demonstrate quantum confinement at the single-layer limit[13]. As with graphene, these layered materials can be easily exfoliated mechanically to provide monolayers[3-7,14-16] and assume a hexagonal honeycomb structure in which the M and X atoms are located at alternating corners of the hexagons. However, unlike graphene, which has a gapless Dirac cone band structure, MX$_2$ has a rather large bandgap, making these materials more versatile as candidates for thin, flexible device applications and useful for a variety of other applications including lubrication[16], catalysis[17], transistors[18] and lithium-ion batteries[19]. Most interestingly, an indirect to direct bandgap transition in the monolayer limit has been predicted theoretically and supported experimentally by optical measurements[3-5,9,12]. Because of the direct bandgap, monolayer MX$_2$ is favourable for optoelectronic applications5 and field-effect transistors[15,16,18]. Furthermore, both the conduction and valence bands have two energy degenerate valleys at corners of the first Brillouin zone, making it viable to optically control the charge carriers in these valleys and suggesting the possibility of valley-based electronic and optoelectronic applications[3,6-8]. Despite these exciting developments, direct experimental verification of the novel band structure at the monolayer limit remains lacking. Furthermore, for many applications, it is vital to manufacture high-quality epitaxial films with controllable methods such as chemical vapour deposition (CVD) or molecular beam epitaxy (MBE)[20,21].

In this Letter, we report layer-by-layer growth of high-quality single-crystal $MoSe_2$ thin films by MBE on an epitaxial grapheneterminated 6H-SiC(0001) substrate[22]. Our in situ angle-resolved photoemission spectroscopic (ARPES) study provides the first direct experimental evidence of the distinct transition in the bandstructure for thin film samples with thicknesses ranging from one monolayer (ML) to eight monolayers. Moreover, we find rather large spin-splitting (~180 meV) at the valence band maximum (VBM) of the monolayer $MoSe_2$ film, a signature of the combined effects of spin-orbit coupling and inversion symmetry breaking.

Figure 1a presents the crystal structure of the layered $MoSe_2$. Each single layer (Se-Mo-Se) of $MoSe_2$ consists of two layers of Se atoms on opposite faces, with one layer of Mo atoms inserted in the middle. Figure 1b shows the reflection high-energy electron diffraction (RHEED) pattern of our substrate of epitaxial bilayer graphene-terminated 6H-SiC(0001)[22]. The similar layered structure and chemically inert surface of graphene make it a perfect substrate for van der Waals epitaxial growth of two-dimensional layered materials[21,23,24]. We have successfully grown high-quality singlecrystal $MoSe_2$ films of large size (~5 mm × 2 mm), from monolayer up to 8 ML, with layer-by-layer control of thickness, by delicate control of the growth conditions. Here, we use the term 'monolayer' to refer to the one-unit-cell triple layer (Se-Mo-Se) of $MoSe_2$ (correspondingly, the terms 'bilayer' and 'trilayer' refer to two and three layers of the Se-Mo-Se structure). Figure 1c presents the RHEED pattern for our MBE-grown monolayer $MoSe_2$ thin film. The disappearance of the graphene pattern (Fig. 1b) and the appearance of a distinct $MoSe_2$ (1 × 1) pattern (Fig. 1c), which is insensitive to both sample (~5 mm × 2 mm) and beam positions, indicate the growth of a homogeneously well-structured film. Figure 1e makes a direct comparison of our calculated band structures (Fig. 1d) and the corresponding ARPES spectra (Fig. 1f) of the monolayer $MoSe_2$ film along the Γ-K direction in the hexagonal Brillouin zone. The contribution from bilayer graphene is not visible in this momentum window, because it centres further in $k_y$, the momentum along the Γ-K direction for both $MoSe_2$ films and graphene, at the K point of the bilayer graphene substrate (Supplementary Figs 1, 2). Despite the energy scale difference, the calculation and ARPES spectra show good qualitative agreement.

Renormalizing the energy scale of calculation by ~17%, we found that the calculated bands (dotted lines in Fig. 1e) are in good agreement with the second-derivative spectra (Fig. 1e). More interestingly, the difference in the relative position of the valence bands at the Γ point and K point in the monolayer film is significantly larger than that obtained from the theory, indicating that $MoSe_2$ shows a stronger tendency towards a direct bandgap material than predicted theoretically.

Figure 2a-d presents the ARPES spectra of monolayer, bilayer, trilayer and 8 ML $MoSe_2$ films, respectively. The second-derivative spectra in Fig. 2e-h are provided to enhance the visibility. Comparisons with our calculations based on the generalized gradient approximation (GGA25; Fig. 2i-l) clearly show that the thickness-dependent band structure evolution is highly consistent with theoretical calculations. In particular, in the spectra from the monolayer $MoSe_2$ film (Fig. 2a,e), the VBM at the K point (-1.53 eV) is significantly higher than the Γ point valence band (-1.91 eV). However, in bilayer and thicker films, the VBM switches to the Γ point (Fig. 2a-h). The quantum confinement effect, which can reveal the number of layers, can be seen around the top valence band at the Γ point. In monolayer $MoSe_2$ film there is only one band above the binding energy of 22 eV at the Γ point (Fig. 2a,e), but in the bilayer film this band evolves into two branches, and then to three branches in the trilayer film (Fig. 2a-h). In the 8 ML film, the calculation (Fig. 2l) shows the presence of eight branches, although in Fig. 2d and h we can only see two main, broad branches due to the limited resolution (these branches are quite close to each other). This significant step-by-step evolution of the valence band provides a straightforward method to identify the thickness of ultrathin $MoSe_2$ films, and also provides verification of the layer-by-layer growth mode of our thin film.

Figure 3a,b presents the second-derivative spectra of undoped and potassium-doped monolayer $MoSe_2$ films, respectively. With surface doping with potassium, we can raise the chemical potential of the $MoSe_2$ film (the green dashed lines in Fig. 3a,b indicate the ~130 meV movement of the valence band). This enabled us to observe how the conduction band minimum (CBM) dropped below the Fermi level in potassium-doped monolayer $MoSe_2$ film (Fig. 3b). Figure 3c,d shows the constant energy maps at the CBM (Fig. 3b) and VBM (Fig. 3a). We can see

that both the CBM and VBM are located at all the K points in the Brillouin zone (no photoemission intensity was observed at the Γ point), which implies the presence of a direct bandgap at the six K points in monolayer MoSe$_2$. In Fig. 3b we measured this direct bandgap to be ~1.58 eV, which is very close to the value of 1.55 eV reported by a photoluminescence experiment in mechanically exfoliated monolayer MoSe$_2$ (ref. 4).

Figure 3e,f shows the second-derivative spectra of undoped and potassium-doped 8 ML MoSe$_2$ films, respectively. With the same doping method, we found that the valence band moved by ~0.46 eV. We can also observe the CBM in the spectra of potassium-doped 8 ML MoSe$_2$ film (Fig. 3f). Figure 3g,h presents the constantenergy maps at the CBM (Fig. 3f) and VBM (Fig. 3e). In contrast to the monolayer MoSe$_2$ film, in the 8 ML MoSe$_2$ film the CBM is still at the six K points (Fig. 3g), but the VBM is at the Γ point (Fig. 3h). Thus, the 8 ML MoSe$_2$ displays an indirect bandgap of ~1.41 eV (Fig. 3f). We note that the calculated GGA bandgap values (Fig. 2i-l) are underestimated, which is a well-known problem in that GGA density functional theory generally underestimates the bandgaps in semiconductors and insulators[9].

From monolayer to 8 ML, we found that the CBM does not change its position at the K point. Our calculations also show that the CBM stays at the K point. However, from monolayer to bilayer and thicker films, both our ARPES spectra and the calculations show that the VBM switches from the K point to the Γ point. Because the interaction between the graphene substrate and the van der Waals epitaxial MoSe$_2$ film is found to be minimal in both experiment (Supplementary Fig. 3) and calculation (Supplementary Fig. 4), this VBM evolution clearly indicates the direct to indirect bandgap transition in going from monolayer to bilayer MoSe$_2$.

Another interesting observation is that we found a very clear band-splitting of the VBM at the K point of monolayer MoSe$_2$ film (Figs 2a,e and 3a). A similar band-splitting can also be seen in bilayer, trilayer and 8 ML MoSe$_2$ films (Fig. 2f-h). Our calculations show that this splitting is mainly controlled by the strength of the spin-orbit coupling. For an odd number of layers, there is no inversion symmetry, and each state at the K point is spin-nondegenerate (Fig. 2i and inset of Fig. 2k, where red and blue round spots indicate different spin polarization). For an even number of layers, inversion symmetry is restored, and every state becomes spin-degenerate (Fig. 2j and inset of Fig. 2l). The combined effects of the spin-orbit coupling and inversion symmetry-breaking can be best seen by a comparison between monolayer and bilayer MoSe$_2$ films. In the monolayer MoSe$_2$ film, each state at the K point is spin-nondegenerate while states at the Γ point are spindegenerate (because K is not a time-reversal-invariant point, but Γ is). The top two branches that start at the K point, merging into one branch at the Γ point, are observed both in the experiment (Fig. 2a,e) and theory (Fig. 2i). Each spin split state is predicted to be nearly 100% spin-polarized (Fig. 2i). Such a high degree of spin polarization has also recently been predicted in silicene thin films in proposals for a high-efficiency spin filter[26]. In bilayer MoSe$_2$ film, the band is doubly degenerate without spin-splitting (Fig. 2j). Both experiment and theory exhibit two branches on top of the valence bands at both the Γ and K points (Fig. 2f,j). In the trilayer MoSe$_2$ film, the magnitude of the spin-splitting within the two main branches is only a few meV, making them nearly degenerate (inset of Fig. 2k), so we can only observe two branches in ARPES in Fig. 2g. In the 8 ML thin film, the calculated eight spin-degenerate states at the K point (inset in Fig. 2l) merge into two blurred branches in the ARPES spectra (Fig. 2h). Our finding of the spin-splitting of ~180 meV in monolayer MoSe$_2$ is consistent with a previous theoretical prediction (183 meV)[10] and larger than that in monolayer MoS2 (as measured recently by triply resonant Raman scattering: ~100 meV)[27]. This spin signature with larger spin-splitting gives the layered MoSe$_2$ greater application potential than MoS2 in spintronic devices, as well as a new basis on which to investigate spin-orbit physics beyond topological insulators[28].

To summarize, we have successfully achieved layer-by-layer growth of high-quality MoSe$_2$ thin films using MBE. The ARPES study shows a distinct transition from an indirect bandgap of ~1.41 eV to a direct bandgap ~1.58 eV when the layer thickness changes from 8 ML to monolayer. Together with the corresponding first-principles computations, this not only provides direct experimental proof of the novel electronic structure evolution, but also reveals clear spin-split bands only in the monolayer limit.

## Methods

**Thin film growth and ARPES.** This experiment was performed at the HERS endstation of beamline 10.0.1, Advanced Light Sources, Lawrence Berkeley National Laboratory. Thin film samples of $MoSe_2$ were grown in the MBE chamber at the beamline with base pressure of $\sim 2 \times 10^{-10}$ Torr, and then transferred directly into the analysis chamber with base pressure of $\sim 3 \times 10^{-11}$ Torr right before the ARPES measurements. Bilayer graphene substrates were prepared by flash annealing of the 6H-SiC(0001) to 1300 $^o$C (ref. 22). . High purity Mo and Se were evaporated from an e-beam evaporator and a standard Knudsen cell, respectively. Flux ratio of Mo to Se was controlled to be about 1:8. Growth process was monitored by *in-situ* RHEED system and growth rate was about 8.5 minutes per monolayer. During the growth process the substrate temperature was kept at 250 $^o$C and after growth the sample was annealed to 600 $^o$C for 30 minutes. The potassium for surface doping was evaporated from a SAES Getters alkali metal dispenser. The ARPES data were taken with a Scienta R4000 electron analyzer. Samples were cooled down to 40 K by liquid helium during measurements. The photon energy was set at 70 eV, with energy and angular resolution of 25 meV and 0.1 degree, respectively. The photon polarization direction was set to be 72$^°$ out of the plane of incidence to get evenly distributed even and odd state signal. The size of beam spot on the sample is roughly 150 micro-meter $\times$ 200 micro-meter. We do not find any change in the observed ARPES spectra when changing beam position around the sample (~5 mm $\times$ 2 mm), which suggest the homogeneity of our sample.

**Electronic structure calculations.** The electronic structures are calculated using the full-potential projected augmented wave method[29] as implemented in the VASP package[30] within the generalized gradient approximation (GGA) scheme[25]. The spin-orbit coupling was included self-consistently, and a 15$\times$15 Monkhorst-Pack k-point mesh was used. The thin films are modeled as slabs separated by vacuum with thickness of about 15 Å.

## References

1. Klitzing, K. v., Dorda, G. & Pepper, M. New method for high-accuracy determination of the fine-structure constant based on quantized hall resistance. *Phys. Rev. Lett.* **45**, 494-497 (1980).
2. Geim, A. K. Graphene: status and prospects. *Science* **324**, 1530-1534 (2009).
3. Mak, K. F. *et al.* Atomically Thin $MoS_2$: A new direct-gap semiconductor. *Phys. Rev. Lett.* **105**, 136805 (2010).
4. Tongay, S. *et al.* Thermally driven crossover from indirect toward direct bandgap in 2d semiconductors: $MoSe_2$ versus $Mos_2$. *Nano Lett.* **12**, 5576-5580 (2012).
5. Splendiani, A. *et al.* Emerging photoluminescence in monolayer $MoS_2$. *Nano Lett.* **10**, 1271-1275 (2010).
6. Cao, T. *et al.* Valley-selective circular dichroism of monolayer molybdenum disulphide. *Nature Commun.* **3**, 887 (2012).
7. Zeng, H. *et al.* Valley polarization in $MoS_2$ monolayers by optical pumping. *Nature Nanotech.* **7**, 490-493 (2012).
8. Xiao, D. *et al.* Coupled spin and valley physics in monolayers of $MoS_2$ and other group-VI dichalcogenides. *Phys. Rev. Lett.* **108**, 196802 (2012).
9. Ellis, J. K., Lucero, M. J. & Scuseria, G. E. The indirect to direct band gap transition in multilayered $MoS_2$ as predicted by screened hybrid density functional theory. *Appl. Phys. Lett.* **99**, 261908 (2011).
10. Zhu, Z. Y., Cheng, Y. C. & Scwingenshlögl, U. Giant spin-orbit-induced spin splitting in two-dimensional transition-metal dichalcogenide semiconductors. *Phys. Rev. B* **84,** 153402 (2011).
11. Cheiwchanchamnangij, T. & Lambrecht, W. R. L. Quasiparticle band structure calculation of monolayer, bilayer, and bulk $MoS_2$. *Phys. Rev. B* **85**, 205302 (2012).
12. Kumar, A. & Ahluwalia, P. K. Electronic structure of transition metal dichalcogenides monolayers 1H-$MX_2$ (M = Mo, W; X = S, Se, Te) from ab-initio theory: new direct band gap semiconductors. *Eur. Phys. J. B* **85**, 186 (2012).
13. Balendhran, S. *et al.* Two-dimensional molybdenum trioxide and dichalcogenides. *Adv. Funct. Mater.* **23**, 3952-3970 (2013).
14. Radisavljevic, B. *et al.* Single-layer $MoS_2$ transistors. *Nature Nanotech.* **6**, 147-150 (2011).
15. Larentis, S., Fallahazad, B. & Tutuc, E. Field-effect transistors and intrinsic mobility in ultra-thin $MoSe_2$ layers. *Appl. Phys. Lett.* **101**, 223104 (2012).

16 Lee, C. *et al.* Frictional characteristics of atomically thin sheets. *Science* **328**, 76-80 (2010).

17 Laursen, A. B., Kegnæs, S., Dahl, S. & Chorkendorff, I. Molybdenum sulfides—efficient and viable materials for electro - and photoelectrocatalytic hydrogen evolution. *Energy Environ. Sci.* **5**, 5577-5591 (2012).

18 Yoon, Y., Ganapathi, K. & Salahuddin, S. How good can monolayer $MoS_2$ transistors be? *Nano Lett.* **11**, 3768-3773 (2011).

19 Chang, K. & Chen, W. In situ synthesis of $MoS_2$/graphene nanosheet composites with extraordinarily high electrochemical performance for lithium ion batteries. *Chemical Commun.* **47**, 4252 (2011).

20 Liu, K.-K. *et al.* Growth of large-area and highly crystalline $MoS_2$ thin layers on insulating substrates. *Nano Lett.* **12**, 1538-1544 (2012).

21 Shi, Y. *et al.* Van der Waals epitaxy of $MoS_2$ layers using graphene as growth templates. *Nano Lett.* **12**, 2784-2791 (2012).

22 Wang, Q. *et al.* Large-scale uniform bilayer graphene prepared by vacuum graphitization of 6H-SiC(0001) substrates. *J. Phys.* **25**, 095002 (2013).

23 Zhang, Y. *et al.* Crossover of the three-dimensional topological insulator $Bi_2Se_3$ to the two-dimensional limit. *Nat. Phys.* **6**, 584-588 (2010).

24 Koma, A. Van der Waals epitaxy—a new epitaxial growth method for a highly lattice-mismatched system. *Thin Solid Films* **216**, 72-76 (1992).

25 Perdew, J. P., Burke, K. & Ernzerhof, M. Generalized gradient approximation made simple. *Phys. Rev. Lett.* **77**, 3865-3868 (1996).

26 Tsai, W.-F. *et al.* Gated silicene as a tunable source of nearly 100% spin-polarized electrons. *Nature Commun.* **4**, 1500 (2013).

27 Sun, L. *et al.* Spin-orbit splitting in single-layer $MoS_2$ revealed by triply resonant raman scattering. *Phys. Rev. Lett.* **111**, 126801 (2013).

28 Hasan, M. Z. & Kane, C. L. Colloquium: topological insulators. *Rev. Mod. Phys.* **82**, 3045 (2010).

29 Blöchl, P. E. Projector augmented-wave method. *Phys. Rev. B* **50**, 17953-17979 (1994).

30 Kresse, G. & Furthmüller, J. Efficient iterative schemes for ab initio total-energy calculations using a plane-wave basis set. *Phys. Rev. B* **54**, 11169-11186 (1996).

**Acknowledgements**

The work at the ALS is supported by the US DoE Office of Basic Energy Science under contract No. DE-AC02-05CH11231. The work at the SIMES and Stanford University is supported by the US DoE Office of Basic Energy Science under contract No. DE-AC02-76SF00515. The work at Oxford University is supported from a DARPA MESO project (No. 187 N66001-11-1-4105). The work at Northeastern University is supported by the US DoE Office of Basic Energy Sciences under contract No. DE-FG02-07ER46352, and benefited from Northeastern University's Advanced Scientific Computation Center (ASCC), theory support at the Advanced Light Source, Berkeley and the allocation of time at the NERSC supercomputing center through DoE grant number DE-AC02-05CH11231. T.R.C. and H.T.J. are supported by the National Science Council, Taiwan. H.T.J also thanks NCHC, CINC-NTU and NCTS, Taiwan for technical support.

**Author contributions**

Y.Z. led the thin film growth effort with F.S., J.L., R.M. and S.K.M, performed ARPES measurements with B.Z., Z.L. and S.K.M., and analyzed the data. Y.Z., H.L. and S.K.M. wrote the paper with suggestions and comments by A.B. and Z.X.S. Y.T.C. and Y.H. characterized samples with Raman and AFM. T.R.C., H.L., H.T.J. and A.B. provided theoretical support. S.K.M., Y.L.C., Z.H., A.B. and Z.X.S. are responsible for project direction, planning and infrastructure.

**Additional Information**

Supplementary information is available in the online version of the paper. Reprints and permissions information is available online at www.nature.com/reprints. Correspondence and requests for materials should be addressed to S.K.M. and Z.X.S.

**Competing financial interests**

The authors declare no competing financial interests.

**Figure 1 Crystal structure, RHEED patterns and overall ARPES spectra of MoSe$_2$ thin film. a** Crystal structure of MoSe$_2$. **b & c** RHEED patterns of **b** epitaxial bilayer graphene over 6H-SiC(0001) substrate and **c** monolayer MoSe$_2$ thin film grown on the substrate. **d** Theoretical band structures calculated using GGA, along the Γ-K direction of the monolayer MoSe$_2$ film. The zero energy represent valence band maximum. **e** Direct comparison of theoretical and experimental band structures of the monolayer MoSe$_2$ film. The experimental band structure is a second derivative of data shown in **f** to enhance the visibility (black and white intensity plot), and the overlaid green dotted lines are the calculated band structures with modified energy gauge. $k_y$ refers to the momentum along the Γ-K direction, corresponding with the y-axis shown in **a**.

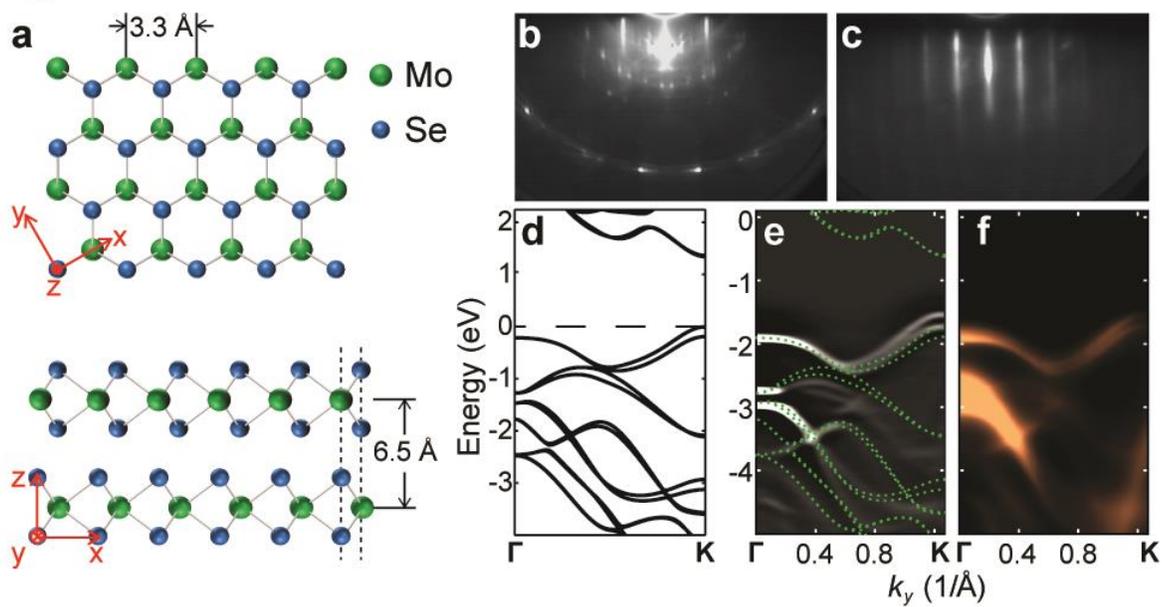

**Figure 2 Band evolution with increasing thickness in MoSe₂ thin films. a-d** ARPES spectra of monolayer, bilayer, trilayer and 8 ML MoSe₂ thin films along Γ-K direction. The white and green dotted lines indicate the energy positions of the apexes of valence bands at the Γ and K points, respectively, with energy values written in the same colours. **e-h** Second derivative spectra of **a-d**, respectively, to enhance the visibility of some bands. The yellow dashed lines are the Fermi level. **i-j** Calculated band structures of monolayer, bilayer, trilayer and 8 ML MoSe₂. Insets show zoom-in splitting of valence band at the K point. Blue and red round spots indicate opposite spin directions.

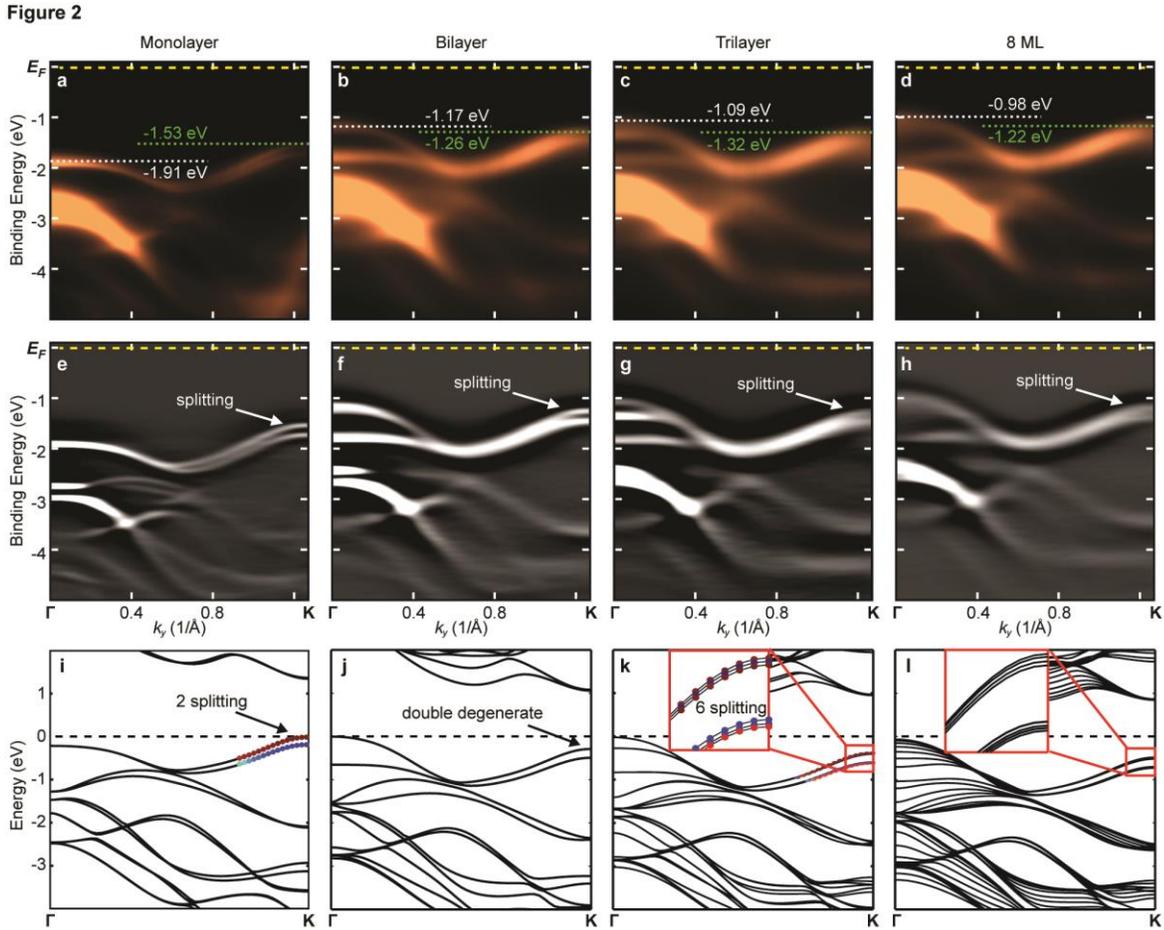

**Figure 3 Direct bandgap in monolayer and indirect bandgap in 8 ML MoSe$_2$ thin films. a-d** ARPES data for monolayer sample. **e-h** ARPES data for 8 ML sample. **a & e** Second derivative spectra of undoped **a** monolayer and **e** 8 ML MoSe$_2$ films along Γ-K direction. **b & f** Second derivative spectra along Γ-K direction after potassium surface doping to shift chemical potential and reveal CBM in monolayer and 8 ML MoSe$_2$ thin films, respectively. The yellow dashed lines are the Fermi levels. Blue and red arrows in **a** are to indicate the opposite spin directions of the spin-split states near the K point in the monolayer MoSe$_2$ film. The green dashed lines indicate the valence bands in monolayer and 8ML MoSe$_2$ films moved by 0.13 eV and 0.46 eV with potassium doping, respectively. **c & g** Constant energy maps at the CBM of potassium doped monolayer and 8 ML MoSe$_2$ films, respectively. **d & h** Constant energy maps at the VBM of undoped monolayer and 8 ML MoSe$_2$ films, respectively. The red hexagons indicate the first Brillouin zone of the system. $k_x$ and $k_y$ refer to the momentum along the Γ-K and Γ-M direction, corresponding with the x-axis and y-axis shown in Fig. 1**a**, respectively.

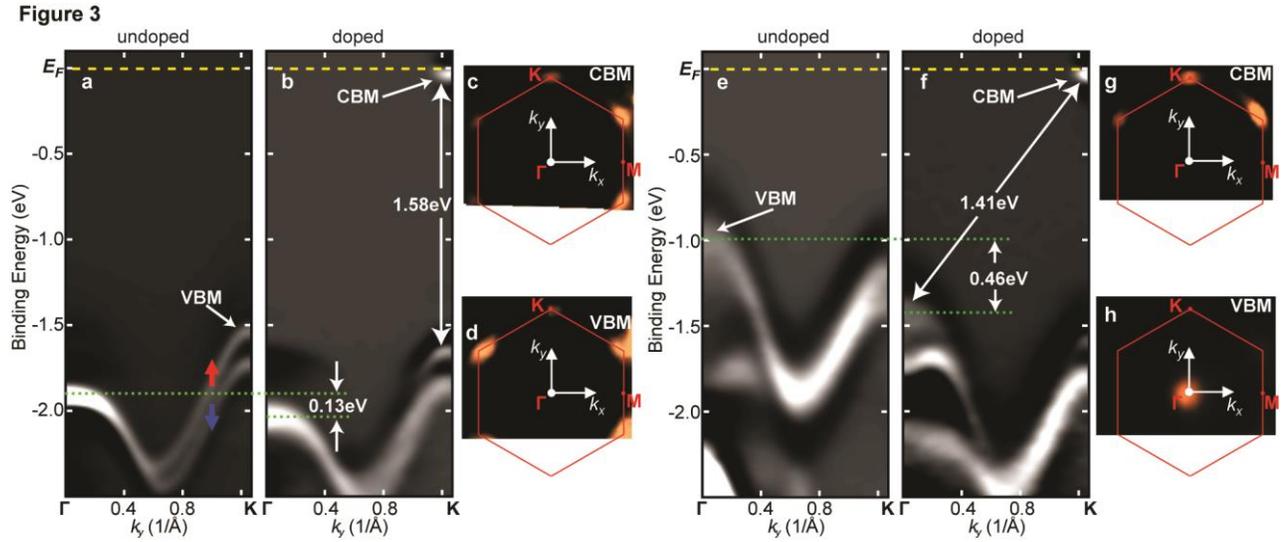